\RequirePackage[2020-02-02]{latexrelease}
\documentclass[prodmode,acmtap]{ci2018}

\acmVolume{2}
\acmNumber{3}
\acmArticle{1}
\articleSeq{1}
\acmYear{2010}
\acmMonth{5}

\usepackage[ruled]{algorithm2e}
\usepackage{outlines}
\usepackage{verbatim}
\usepackage{booktabs}
\usepackage{dirtytalk}
\usepackage[normalem]{ulem}
\usepackage{pdfpages}
\usepackage{graphicx} 

\SetAlFnt{\algofont}
\SetAlCapFnt{\algofont}
\SetAlCapNameFnt{\algofont}
\SetAlCapHSkip{0pt}
\IncMargin{-\parindent}
\renewcommand{\algorithmcfname}{ALGORITHM}

\title{Designing Sousveillance Tools for Gig Workers}

\author{Maya De Los Santos*, Kimberly Do*, Michael Muller, Saiph Savage\\Northeastern University, IBM Research, Universidad Nacional Autonoma de Mexico (UNAM)\\ \** Both authors contribuited equally to the research.}

\begin{document}
\maketitle

\begin{figure}[h]
    \centering
    \includegraphics[width=1.00\textwidth,scale=1.0]{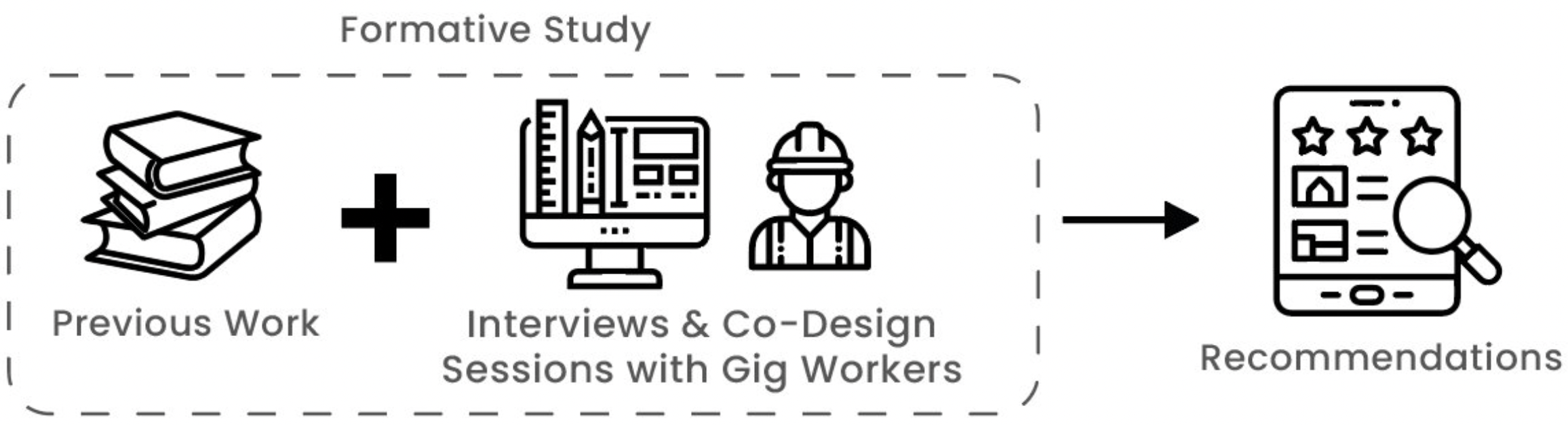}
    \par{Overview of our process for creating design recommendations for sousveillance tools for gig workers.}
    \label{fig:teaser}
\end{figure}

\section{Abstract}
As independently-contracted employees, gig workers disproportionately suffer the consequences of workplace surveillance, which include increased pressures to work, breaches of privacy, and decreased digital autonomy. Despite the negative impacts of workplace surveillance, gig workers lack the tools, strategies, and workplace social support to protect themselves against these harms. Meanwhile, some critical theorists have proposed sousveillance as a potential means of countering such abuses of power, whereby those under surveillance monitor those in positions of authority (e.g., gig workers collect data about requesters/platforms). To understand the benefits of sousveillance systems in the gig economy, we conducted semi-structured interviews and led {co-design} activities with gig workers.  {We use ``care ethics'' as a guiding concept to understand our interview and co-design data, while also focusing on empathic sousveillance technology design recommendations.} Through our study we identify gig workers' attitudes towards and past experiences with sousveillance. {We also uncover the type of sousveillance technologies imagined by workers, provide design recommendations, and finish by discussing how to create empowering, empathic spaces on gig platforms.

\section{Introduction}
The digitization of work has enabled and popularized work arrangements such as gig work, defined as independently contracted work opportunities organized by online gig work platforms \cite{posada2022coloniality}. While the online format of gig work offers flexible work options, it also exposes gig workers to the many privacy risks and constant surveillance associated with online work \cite{sannon2022surveillance}. Consequently, many have raised concerns over the invasive surveillance practices encouraged by gig work platforms \cite{sannon2022surveillance,request2023,newlands2021algorithm}. For example, to monitor and manage their workers, gig work platforms and gig requesters employ ``bossware" such as location trackers, screen recorders, mouse trackers, webcam activators, and keystroke recorders \cite{sannon2022surveillance}. Under the guise of “productivity-enhancing tools,” bossware is commonly used against the best interest of gig workers \cite{kantor2022,mitpress_retailers_2023}. This oftentimes hidden surveillance worsens data imbalances and reduces workers' digital autonomy, sometimes leading to unwarranted productivity claims by requesters \cite{zhang2022}. Consequently, gig work surveillance not only infringes on workers' privacy but also substantially decreases their digital autonomy.

Gig work surveillance also carries significant privacy and health implications, even prompting policy changes. For example, Uber faced scrutiny for unjustly terminating drivers due to tracking algorithms \cite{degeurin2021}. Deliveroo was fined by Italian authorities for violating GDPR transparency and privacy laws with its monitoring algorithms \cite{wie2021}. In India, local gig work platforms were investigated for invasive tracking practices \cite{udhayakumar2022}. Furthermore, gig work surveillance is associated with long-term negative health outcomes, including chronic work-related stress, higher rates of depression, and work anxiety \cite{smith1992,roemmich2023}.

In response to these controversies, gig workers, practitioners, and researchers have begun exploring various approaches to combat surveillance in the gig economy \cite{calacci2022,irani2013turkopticon,salehi2015}. One such approach is through sousveillance---the act of subordinates monitoring people in power \cite{mann2004sousveillance}---which could provide an avenue for workers to fight back against the harms and consequences of surveillance \cite{calacci2022}. According to critical theorist Foucault, by scrutinizing and recording those in positions of authority, sousveillance can: 1) promote organizational transparency; 2) strengthen accountability; and 3) increase feelings of empowerment among subordinates in organizations with high power imbalances \cite{foucault1997}. Drawing inspiration from Foucault's theory, some academics and practitioners have created digital sousveillance tools which provide workers with valuable insights about their requesters and their surveillance. For example, Uber and Lyft drivers have created sousveillance tools to identify and track the payment discrepancies between what the gig platforms charge requesters and what they actually pay the drivers \cite{griffith2022}. Workers and researchers have also utilized digital sousveillance tools to call attention to issues like workers earning less than minimum wage on Amazon Mechanical Turk \cite{hara2018data}, and the burden of unpaid labor on gig platforms \cite{toxtli2021quantifying,li2022all}. When used responsibly, sousveillance tools can empower workers to reveal and retaliate against hidden injustices in the gig economy.  

However, despite recent efforts to develop digital sousveillance tools, two persistent challenges arise. First, existing tools lack human-centered design (HCD) practices, failing to align optimally with workers' needs and work practices. Second, they have not effectively addressed the social intricacies of sousveillance in the gig economy \cite{lee2021}. Workers must exercise caution when using such tools to avoid disrupting relationships with requesters, violating gig platform policies, or risking job loss. Therefore, HCD is crucial for developing gig sousveillance tools that align with workers' expectations and experiences \cite{zhang2023,hsieh}. Without an HCD perspective, we risk creating tools that do not match gig workers' vision, preferences, and personal needs \cite{lee2021,zytko2022}.  {However, to effectively design sousveillance technologies for gig work, it is essential to account for the intricate social dynamics involved, as previously mentioned. We thus use care ethics as our guiding framework (sensitizing concept) \cite{nel1984feminine,glaser2017discovery}. Care ethics helps us to focus on the interdependence in relationships among gig workers, technology, and stakeholders in sousveillance \cite{engster2011care}, as well as highlight the impact of these dynamics on sousveillance practices, while addressing the vulnerabilities and power imbalances in gig work \cite{reynolds2016infinite}. Care ethics offers a framework for designing sousveillance tools that prioritize the wellbeing of workers \cite{hamington2019integrating}.} {Based on these ideas}, we conducted interviews and {co-design} sessions with gig workers to inform the creation of worker-centric gig sousveillance tools. 
Care ethics shaped our interview protocol, design activities, data analysis, and design recommendations. 

Our research is thus driven by the following research questions:

\begin{itemize}
  \item RQ1: How do gig workers understand and conceptualize sousveillance on gig platforms?
  \item RQ2: What are gig workers’ attitudes towards sousveillance?
  \item RQ3: What are the risks that gig workers may face when conducting sousveillance?
  \item RQ4: How do gig workers currently conduct sousveillance, if at all?
   \item RQ5: How do gig workers envision the future of sousveillance tools?
  \item {RQ6: How can care ethics enhance our understanding of gig workers' experiences with sousveillance, and what implications does this have for the future development of sousveillance tools?}
\end{itemize}

{Our study, combining co-design sessions and interviews with gig workers, leads in creating worker-centered sousveillance technologies, emphasizing care ethics.} We observed that all our participants had previous experience with some form of sousveillance, which they viewed as a valuable tool when used responsibly. Sousveillance helped improve labor outcomes, anticipate requesters' needs, and stay competitive in the ever-changing gig economy. Despite recognizing the benefits of gig sousveillance, workers were concerned about associated overhead costs, privacy issues, platform policies, and accommodating requester preferences. Interestingly, sousveillance often strengthened trust between gig workers and requesters by enhancing communication, creating long-term opportunities, and improving rapport. In our {co-design} sessions, workers imagined sousveillance tools that address transparency issues and data imbalances in gig platforms while ensuring data reliability. They also envisioned tools that support their emotional well-being in gig work without belittling the invisible labor they perform. Based on our findings, {we propose design recommendations for sousveillance tools centered on gig worker needs and their care. We conclude by discussing the key findings of our study, as well as discuss the complexities of gig platforms. Despite inherent collaborative features, these platforms often have design elements that impede empathic collaboration among stakeholders. We discuss how worker-centric tools, specifically worker-owned sousveillance technologies grounded in care ethics, can mitigate these issues and cultivate a more compassionate and collaborative gig economy.}

\section{Terminology}
The concept of sousveillance originated from the influential work of critical theorist Foucault, who analyzed power dynamics and collective action within the prison system \cite{foucault1997}. Foucault advocated for exposing issues created by those in power and holding them accountable through a reversal of surveillance dynamics, placing observation and documentation in the hands of ordinary individuals \cite{foucault1997}. Steve Mann later officially termed this concept ``sousveillance" \cite{mann2004sousveillance}. Sousveillance could create public awareness of potential misconduct or mistreatment of the surveilled, reducing surveillance intensity and generating public records. For instance, it could balance power dynamics in situations with clear imbalances, such as police searches, government interactions, or medical consultations \cite{mann2002,mann2004sousveillance}.
In this paper, we extend the term ``sousveillance" to encompass any form of data collection or monitoring of clients (referred to as ``requesters") or platforms by gig workers \cite{mann2003sousveillance}.

\section{Related Work}

\subsection{Resisting Workplace Surveillance}
As digital productivity trackers become increasingly pervasive, workers have continuously sought various tools and tactics to lessen the invasiveness of workplace surveillance. For example, in digital labor platform surveys, workers commonly report using encryption tools like VPNs, obscuring or withholding personal identifying information, turning off or physically covering web cameras, or downloading anti-tracking software \cite{sannon2022surveillance}. However, although these tools and techniques can help workers obscure some of their personal data, they fail to address the broader disparities that result from surveillance. To begin with, one of the most significant consequences of gig surveillance is data asymmetry, a phenomenon where workers are neither able to obtain access to their data nor reap the benefits that such data could provide them (e.g., scheduling suggestions, productivity insights, requester contact information) \cite{zhang2022}. Consequently, in situations where requesters use collected performance data to wrongly incriminate their employees, workers often lack the data autonomy to defend themselves with official, un-tampered evidence \cite{whitehouse-pdf,cfpb-pdf}. Secondly, requesters often coerce workers into accepting work contracts that strictly prohibit the use of such ``surveillance-blocking" tools, leaving workers with little choice but to be surveilled. In these situations, such tools unfortunately cannot support workers negotiating with their requesters' terms and conditions. Ultimately, current anti-surveillance solutions neither afford workers data autonomy nor digital freedom of choice. To this end, researchers and practitioners have begun to explore {various digital interventions} to counteract the negative outcomes of surveillance.

{\subsubsection{Worker-Centric Data-Driven Tools}
Although advances in AI technology have enabled work environments that algorithmically capture, monitor, and analyze worker data \cite{davenport2022working,duke2023ai,miller2021futureofowork}; innovative data-driven tools are providing workers with new means to collaborate and collectively challenge the injustices they encounter in the workplace \cite{DigitalWorkerInquiry2023,toxtli2023designing}. For example, the NGO Coworker.org developed a calculator that uses data to clarify worker payments, aiding their advocacy for transparency and fairness amid the rollout of obscure payment algorithms \cite{ShiptTransparencyCalculator}. Other data-driven tools have helped workers to identify how much money their employer is potentially stealing from them \cite{toxtli2021quantifying,calacci2022,WageTheftIntro2023}, measure unpaid time \cite{platform2020wage}, and ensure worker safety \cite{alimahomed2021surveilling,ShromaMonitoring2023}. Through these tools, workers are recognizing the value of data in exposing and combating workplace injustices \cite{toxtli2021quantifying}, leading to increased use of sousveillance technologies \cite{DriversSeatCoop,TheTimeProject,toxtli2021quantifying}. However, the alignment of these tools with workers' needs remains uncertain \cite{gallagher2023digital}. This paper focuses on interviewing workers and engaging them in design activities to develop design directions for the future of worker-centric sousveillance technologies.} 


\subsubsection{Reclaiming the Outcomes of Workplace Surveillance using Sousveillance} 
Inspired by the formative theories of Steve Mann, workplace sousveillance aims to address the limitations of current anti-surveillance methods \cite{mann2013new}. While some past examples of workplace sousveillance include audio-recording work conversations, video-recording unsafe work conditions, or printing workplace emails for personal records, today's digital sousveillance commonly utilizes pop-up blockers, screen recorders, time productivity trackers, and other increasingly technical tools  \cite{sannon2022surveillance}. Unlike traditional anti-surveillance methods, which seek to intervene in the act of surveillance, workplace sousveillance aims to document the format, measure the frequency, and identify the harms of surveillance \cite{taylor2021,cecchinato2021self}. Moreover, the information obtained through sousveillance can empower workers in promoting collective action initiatives that may defy and dissolve unjustified surveillance \cite{washington2021,wu2022reasonable}. For instance, in 2023, two Amazon workers rallied fellow employees to conduct a large-scale sousveillance study that investigated Amazon's employee surveillance \cite{hall2022}. Shortly after workers published their sousveillance findings, Minnesota state lawmakers in the United States enacted policies that mandated large companies like Amazon to provide transparency around \emph{{quotas and work speed metrics used to evaluate workers’ performance}} \cite{hall2022}. Overall, sousveillance is a powerful tool for workers to combat workplace surveillance.

\subsubsection{Adapting Sousveillance for the Gig Economy} 
Similarly, gig work sousveillance also has the potential to reveal the harms of workplace surveillance and to empower workers to speak up against it \cite{wu2022reasonable}. Informally, workers use sousveillance methods such as taking ``progress report" screenshots, recording dash cam footage, and saving chatlogs with requesters \cite{sannon2022surveillance}. However, recent interest in gig work sousveillance has also compelled others to create more formal systems for conducting sousveillance \cite{griffith_2022,salehi2015}. For example, workers at Uber created Para, a sousveillance tool that documents gig drivers' work hours to create a secondary source of accountability \cite{griffith_2022}. Similarly, researchers collaborated with Amazon Turk workers to create We Are Dynamo, a tool that crowd-sources workers' reviews of requesters \cite{salehi2015}. Together, these tools have been shown to alleviate the detriments of surveillance and provide benefits such as reducing wage theft and highlighting workers' unpaid, invisible labor  \cite{griffith_2022,salehi2015}. Nonetheless, gig work sousveillance presents unique challenges compared to traditional workplace sousveillance. In contrast to ``traditional" employees who typically work together in the same physical workplaces, gig workers tend to work remotely and in complete physical isolation from each other \cite{yao2021}. Consequently, the isolation inherent in gig work can lead to differences in the needed formats of sousveillance. Moreover, as independently contracted workers, gig workers are often subjected to the variable, unclear, and irrational policies of their requesters \cite{Wassom_2022}. Lastly, variation among workers may make it difficult for workers to collectively unite and act upon the findings of their individual sousveillance \cite{wu2022reasonable,yao2021}. Altogether, the varying types of gig work and nuances among workers exemplify a need for worker-centric sousveillance tools to better address the individual and variable needs of gig workers. Thus, while prior work includes gig work sousveillance tools that address issues such as wage theft and unpaid labor, our work seeks to investigate how human-centered design methods can be applied to the creation of gig sousveillance tools that address gig worker data autonomy, data asymmetry and digital freedom.

\subsection{Sensitizing Concepts}

 {In order to be paid, gig workers must meet the needs of their requesters. Meeting other people's needs is a form of care, and theorists of ethics of care note that care can be a reciprocal relationship \cite{tronto2013}. Gig work also involves aspects of computer-mediated collaboration and coordination, to clarify the requirements and to report progress \cite{jarrahi2020platformic}. We therefore looked to the research literatures on ethics of care and on collaboration, as sensitizing concepts that could inform our analysis.}  {Sensitizing concepts, as proposed by Bowen \cite{bowen2006grounded}, have been adopted in qualitative analysis. These concepts are not \textit{theories} to be proven or disproven. Rather, they provide guidance for the active work of constructing an interpretation of the data. In this way, Ribes argued that sensitizing concepts suggest ``where to look but not what to see'' \cite{ribes2017notes}. Sensitizing concepts were proposed for grounded theory, where they can suggest (but not constrain) ways of approaching the analysis \cite{bowen2006grounded}. In Glaserian grounded theory, \textit{coding families} can accomplish a similar kind of guidance \cite{glaser2005staying}). Sensitizing concepts have also been used in thematic analysis \cite{georgousis2021teaching,petersson2023case}, which is the analytic method used in this paper.} {In particular, we use care ethics as a sensitizing concept in our study and design of sousveillance technologies for gig workers, recognizing the precarious nature and lack of support in gig work. Using it as a sensitizing concept helps us focus on critical issues and create empathetic, responsive technological solutions. Care ethics also helps us to address power imbalances in gig work. Next, we delve deeper into the concept of care ethics. }

\subsubsection{Ethics of Care}
 {We use Ethics of Care as a sensitizing concept. Ethics of care were proposed by Gilligan \cite{gilligan1993different} as part of a relational morality.} Also known as care ethics (CE),  {ethics of care are a collection of approaches} 
emphasizing interpersonal relationships, empathy, and caring for others in moral decision-making \cite{tronto2013}. CE differs from other normative theories by encouraging individuals to reflect on their connections and responsibilities to others \cite{engster2011care,held2006ethics}, aiming to maximize care. Traditionally, CE has 
 {provided descriptive and interpretive insights}
in healthcare settings like nursing homes, mental health clinics, and hospitals \cite{krause2017care}. More recently, HCI researchers have used CE to explain community-based learning interactions in settings like hackathons and makerspaces \cite{cheong2021,toombs2015}  {and disaster-response and community food-assistance politics (for review, see \cite{boone2023data})}. While CE is sometimes seen as  {being} primarily  {concerned with} 
relational contexts due to its feminist ethics  {origins \cite{gilligan1993different} and} influences \cite{larrabee2016ethic}, scholars like Held and Tronto argue for its application in social and political contexts \cite{tronto2013,held2006ethics}. In this interpretation, CE  {may} provide insights into how businesses should support employees and stakeholders \cite{sep-feminism-ethics}. We anticipate that some of these business-related concepts should apply in gig work as well. Tronto, a leading figure in CE, identifies four care principles: attentiveness, responsibility, competence, and responsiveness \cite{tronto2013}. These principles encompass acknowledging others' needs, taking care of them, being effective in caregiving, and showing empathy for their needs. Tronto also outlines four phases of care: caring about, taking care of, caregiving, and care receiving \cite{tronto2013}. Fig. \ref{fig:ethicscare} provides an overview of  {Tronto's} care ethics and core principles.  {Held \cite{held2006ethics} describes seven care principles. Three of these principles overlap with Tronto's principles \cite{tronto2013}: attentiveness, responsiveness, and taking responsibility. Held proposed four additional care principles: empathy, mutual concern, sensitivity, and trustworthiness \cite{held2006ethics}. Engster added the care principle of respect \cite{engster2011care}. The combined set of principles is thereby attentiveness, competence, empathy, mutual concern, respect, responsiveness, sensitivity, taking responsibility, and trustworthiness.}  {Following on the ideas of Tronto \cite{tronto2013} and Held \cite{held2006ethics},} 
managerial and industrial/organizational psychologists have recently shown interest in Ethics of Care (EC) \cite{ripamonti2021care,antoni2020caring,ley2023care}. Drawing parallels between familial and workplace relationships, some have coined the term ``workplace ethics of care" (WEC) \cite{antoni2020caring}. Unlike EC, WEC centers on relationships among management, employees, and organizations. Workplace care entails actions like fostering career growth, ensuring safety, and promoting team harmony \cite{ley2023care}. While previous discussions on worker well-being and privacy aimed to create tools for efficient work \cite{toxtli2021quantifying}, WEC fosters nuanced discourse about workplace relationships, tensions, and responsibilities without solely focusing on labor outcomes \cite{obrien2013}. Our research employs WEC to design a worker-centric sousveillance system.  {The work of Engster \cite{engster2007heart}, Held \cite{held2006ethics}, and Tronto \cite{tronto2013,tronto2013caring}, summarized above, suggests nine EC attributes to look for as sensitizing concepts in gig work platforms.}

\begin{figure}
    \centering
    \includegraphics[width=400pt]{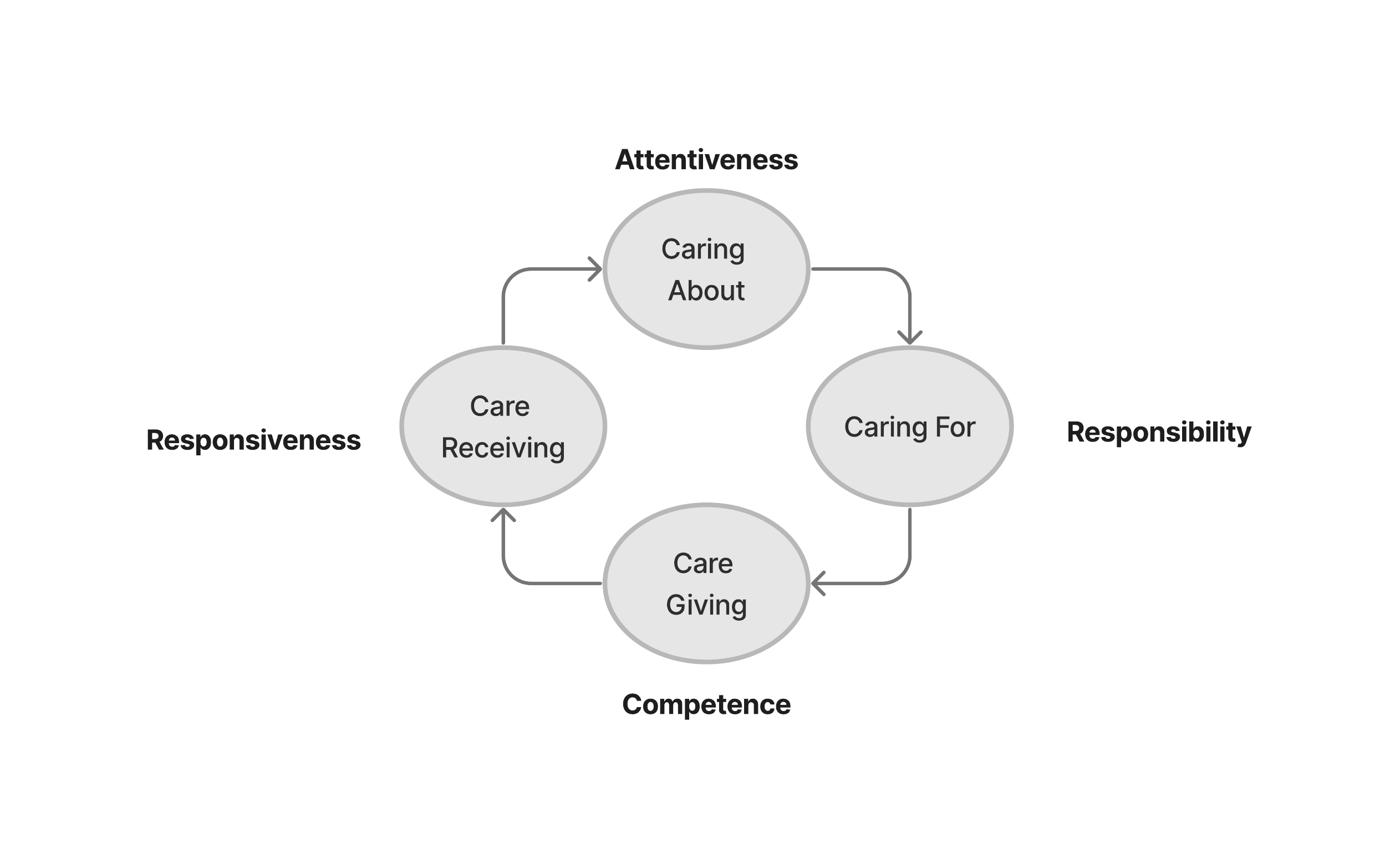}
    \caption{The Ethics of Care---also known as Care Ethics (CE)---as summarized by Tronto, consists of four care principles: attentiveness, responsibility, competence, and responsiveness \cite{tronto2013}. These four principles correspond with four phases of care: caring about, taking care of, caregiving, and care receiving \cite{tronto2013}. Together, these four principles exist in a cycle, where attentiveness leads to responsibility, which leads to competence, and responsiveness, ad infinitum. }
    \label{fig:ethicscare}
\end{figure}






\color{black}
\subsubsection{Collaboration}


Two viewpoints have shaped discourse on gig work systems in recent years. On one hand, gig work systems may be seen as a form of platform-mediated collaboration \cite{boone2023data} in which the digital system(s) facilitate ``a connected yet distanced collective of stakeholders'' \cite{shankar2021coordinating}. Thus, Bj{\o}rn et al. concluded that while collaborative technologies reduced the need for co-presence in many circumstances, they also increased the effort required for \textit{articulation work} (i.e., the invisible work that is often required to coordinate visible ``productive'' work) \cite{strauss1988articulation,strauss1986social,bjorn2014does}. On the other hand, Gherardi and Rodeschini further elaborated the concept of articulation work as a component of ethics of care as performed by organizations \cite{gherardi2016caring}, converging with other accounts of articulation work and ethics of care in nursing \cite{bjornsdottir2018try,lakshmi2022warm,ming2023go}, and more generally the association of articulation work and care work \cite{christensen2011challenges,ertner2019enchanting,lydahl2017visible}. 


In summary, the CSCW literature suggests several topics to look for as sensitizing concepts when we approach gig workers' experiences in this study. Gig work platforms \textit{may} support certain forms of collaboration and coordination; however, they may also \textit{impede} certain human-interaction tasks such as articulation work and relationship work.

\color{black}

\section{Methods}
We undertook an IRB-approved research project composed of semi-structured interviews and {co-design} activities with 16 gig workers from Upwork (n=16). The primary objective of our research was to gain insight into the attitudes, practices, and obstacles experienced by these workers when engaging in sousveillance.  {Note that we also synthesized interview and co-design session outcomes to establish design recommendations for future surveillance technologies for gig workers, integrating care ethics to interpret results and inform these recommendations} \cite{engster2011care,ley2023care}.

\subsection{Participants}
We recruited participants using the popular gig work platform Upwork, which features workers with over 10,000 available skill sets. To reach potential candidates, we posted a job on Upwork inviting workers for our interview study and {co-design} activity \cite{upwork-press-release}. Similar to prior work \cite{fang2022perceptions}, we included all individuals who were at least 18 years old, fluent in written and oral English and had at least 1 year of gig work experience. Our final sample includes 16 gig workers (5 male, 9 female, and 2 non-binary, see Table \ref{tab:participants}). All participants had at least 2 years of gig work experience and completed at least 1 hour of gig work per week. Participants ranged from ages 18 to 38 and resided in the Americas, Africa, and Asia. Our participants did gig work in various areas, including manuscript editing, transcription services, user testing, customer service, content writing, digital marketing, career coaching, and data entry. 

\begin{table}[ht]
\begin{tabular}{p{.5cm} p{.5cm}p{1.5cm}p{2.2cm}p{2.2cm}p{2.2cm}p{3.5cm}}
ID & Age & Gender    & Race                                                                              & Gig Work Tenure & Hours Per Week & Area of Expertise                                                                                                                                                                                        \\ \midrule
P1                 & 24  & Male      & Black or African American                                                      & 2 Years                            & 5 to 10                                  & Fact checking and search evaluation                                                                                                                                                                      \\
P2                 & 25  & Female    & Asian                                                                             & 5 Years                             & 10 to 20                                 & HR, Web Development, Digital Marketing, Search engine optimization (SEO)                                                                                                                                                              \\
P3                 & 30  & Female    & Asian                                                                             & 4 Years                             & 40 or More                               & Copy writing, Script writing, Digital marketing, Email management                                                                                                                                          \\
P4                 & 22  & Female    & Black or African American                                                      & 4 Years                             & 20 to 40                                 & Virtual Assistance                                                                                                                                                                                       \\
P5                 & 23  & Female    & Black or African American                                                   & 5 Years                             & 5 to 10                                  & Content Writing, Copywriting, Content Marketing and Search engine optimization (SEO)                                                                                              \\
P6                 & 30  & Female    & Asian                                                                             & 5 Years                             & 40 or More                               & Customer service                                                                                                      \\
P7                 & 27  & Female    & White                                                                             & 5 Years                             & 5 to 10                                  & Content writing                                                                                                                    \\
P8                 & 38  & Male      & White                                                                             & 4 Years                             & 1 to 5                                   & Career coaching, HR work                                                                                                                                         \\
P9                 & 19  & Male      & Black or African American                                                       & 5 Years                             & 20 to 40                                 & Content writing, Copywriting                                        \\
P10               & 30  & Female    & Black or African American                                                 & 5 Years                             & 1 to 5                                   & Data entry and survey                                                                                                                                                                                    \\
P11                & 28  & Female    & Black or African American                                                 & 3 Years                             & 10 to 20                                 & General administrative tasks                                                                                                                                                                             \\
P12                & 25  & Male      & Asian                                                                             & 2.5 Years                           & 40 or More                               & Content writing, HR Work                                                                                                                                                                             \\
P13               & 18  & Female    & Asian                                                                             & 2 Years                             & 5 to 10                                  & Content Writing                                                                                                                                 \\
P14                & 28  & Nonbinary & Black or African American, White, Native American; Hispanic & 8 Years                             & 40 Hours or More                               & Professional Development, Content writing, User testing, HR work, Transcription \\
P15                & 25  & Nonbinary & Latino; Hispanic                                                                  & 5 Years                             & 5 to 10                                   & Content writing                                                                                                                                            \\
P16                & 40  & Male & Black or African American                                                                 & 10 Years                             & 40 or More                                   &                         Research, Content writing
                           
\end{tabular}
\caption{}
\label{tab:participants}
\end{table}

\subsection{Interview Protocol}
Our interview protocol centers around our five research questions. More specifically, we authored questions to understand gig workers' attitudes, practices, and barriers towards conducting sousveillance. We prefaced our interviews by asking workers if they had previously heard of the term ``sousveillance" and offered a simple definition if they were unfamiliar. We then asked participants about 1) what they believe sousveillance looks like in practice, 2) whether or not they conduct sousveillance, 3) what methods they use to conduct sousveillance, 4) what barriers they might face when conducting sousveillance, 5) how they feel about the idea of conducting sousveillance, and 6) in what way sousveillance might benefit their gig work. As certain themes emerged during the interviews, we followed up with additional questions to probe workers about what kinds of data they would collect from requesters, how sousveillance could impact their gig work relationships, and reasons why they would feel uncomfortable conducting sousveillance.

We conducted and recorded all interviews remotely using the Zoom online video conference software. Interviews were then transcribed using Otter.ai. We conducted interviews in parallel with data analysis and recruited participants until we achieved data saturation in our thematic analysis. Each interview was between 30 minutes to 45 minutes. 


\subsection{Co-Design Activity}

\begin{figure}
    \centering
\includegraphics[width=330pt]{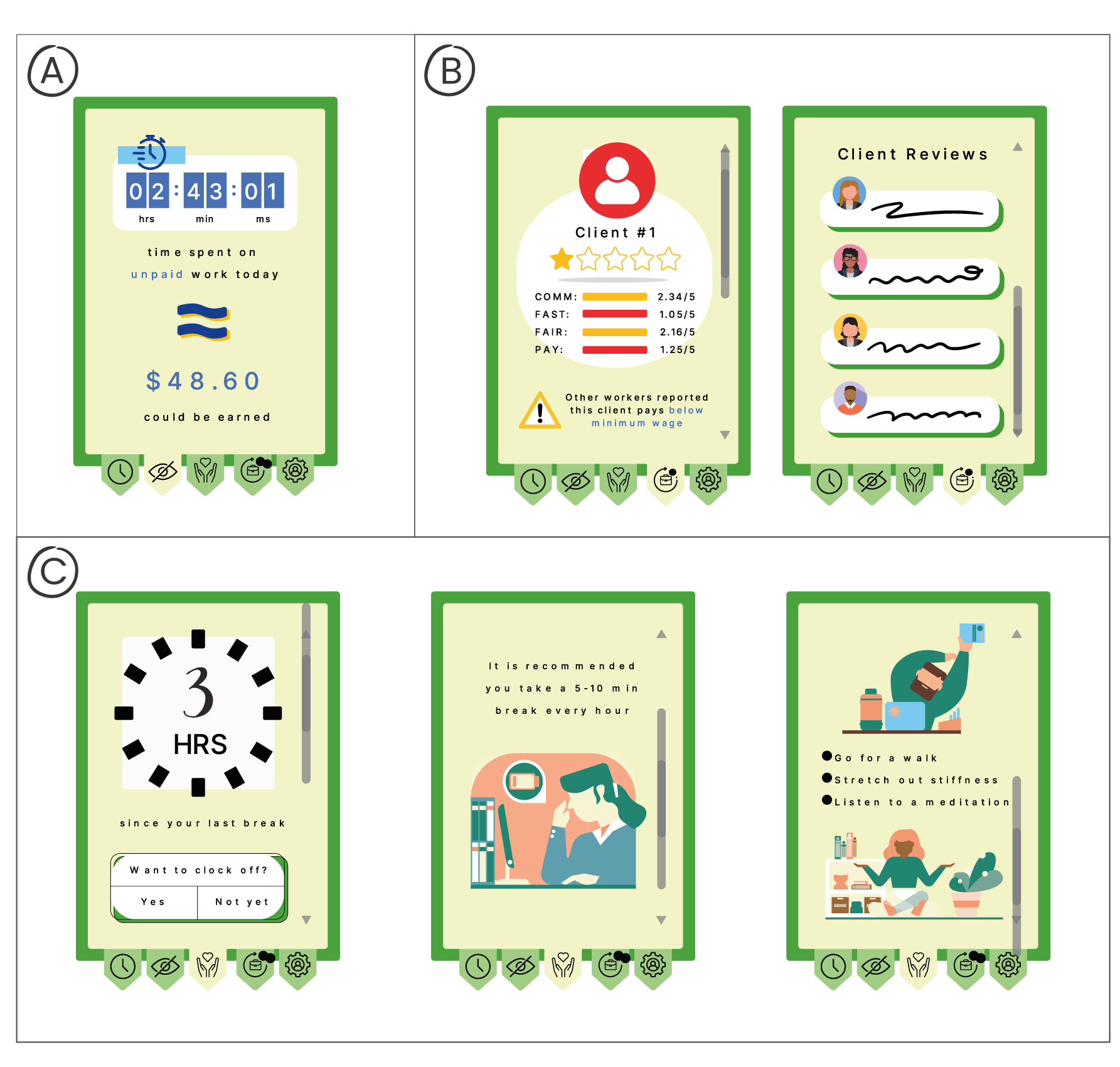}
    \caption{\textbf{Gig Sousveillance Tool Mockups.} We shared three design mockups during our  {co-design} activity to probe participants on their vision for gig sousveillance tools. These mockups include A) a labor tracking tool, B) a requester review dashboard, and C) a wellness suggestions and reminder tool.}
    \label{pd-probe}
\end{figure}

After conducting each interview, we organized a  {co-design} activity where participants had the opportunity to interact with three digital gig work tool mockups---a well-being tracker, a requester review board, and an invisible labor tracker. Our primary objective was to gather crucial insights, feedback, and suggestions from the workers. This valuable input would inform and guide the design of future gig sousveillance tools, ensuring that they effectively cater to the needs and preferences of the workers themselves. These mockups were designed to capture different dimensions of participants' thoughts, emotions, and perceptions, encouraging them to reflect upon and express their opinions freely. The presentation of the mockups was accompanied by open-ended questions that facilitated the participants' engagement. The mockups we developed were built upon prior research that started to delve into tools enabling workers to monitor their workplaces and obtain personalized data for their own purposes. Our focus revolved around three main mockups, each targeting specific aspects: A) documenting the extent of unpaid labor \cite{toxtli2021quantifying}; B) documenting and reviewing interactions with requesters \cite{salehi2015} C) documenting workplace interactions to promote worker well-being. Fig \ref{pd-probe} includes images of the three mockups we explored in the  {co-design} sessions. See Supplementary Materials for the complete list of interview questions and   {co-design} activity workbook.
\subsection{Analysis}

\subsubsection{Interview Analysis}
We qualitatively analyzed all interviews using a inductive, bottom-up thematic analysis. First, two researchers independently reviewed half of all the transcripts to develop two initial codebooks. Then, the researchers reviewed and open-coded the remaining interviews together to iteratively refine a single codebook. The final codebook consisted of 6 top-level codes encompassing platform hierarchies, current approaches, desired outcomes, perceptions, barriers, and risks of sousveillance. Finally, the researchers reconvened to develop themes based on the final codebook using affinity diagramming. We conclude with four top-level themes (see Section \ref{ResultsInterview}).

\subsubsection{  {Co-design} Activity Analysis}
We also qualitatively analyzed all  {co-design} activities using a inductive, bottom-up thematic analysis. First, two researchers independently reviewed half of all the transcripts to develop two initial codebooks. Then, the researchers reviewed and open-coded the remaining transcripts together to iteratively refine a single codebook. The final codebook consisted of 3 top-level codes highlighting the attitudes, potential use cases, and concerns workers expressed about the mockups during the activity. Next, the researchers utilized these codes to inform a list of potential feature improvements and additions.

By utilizing data mockups in this  {co-design} activity, we aimed to gather rich qualitative data that provided valuable insights into the participants' perspectives, preferences, and needs. The valuable input we received from the gig workers directly informed subsequent design iterations of the sousveillance tool mockups. We used the results of our interviews and co-design activities to inform design recommendations for sousveillance technologies tailored for gig workers.

\section{Results: Interview Study}\label{ResultsInterview}
Our analysis surfaced four high-level themes: (1) gig workers' general attitudes towards sousveillance; (2) the potential benefits and challenges that sousveillance can offer gig workers; (3) the conditions and mechanisms necessary to support sousveillance in practice; and (4) finally, how sousveillance transforms existing freelancer-requester relationships. Participants are identified with a "P", followed by their participant number. All quotes have been anonymized and de-identified to protect participant privacy.  {Sensitizing concepts, where relevant, are indicated in italics within angle-brackets (e.g. ``\emph{$<$empathy$>$}'').}

\subsection{Strategies for Conducting Sousveillance.}
When asked if they had heard of the term ``sousveillance", all 16 gig workers initially associated the word with ``surveillance" and did not know what exactly sousveillance was. Despite this unfamiliarity, we discovered that each of them had engaged in sousveillance during their gig work experiences. Our interviews unveiled three main strategies by which gig workers actively practice sousveillance: (1) Conducting online searches for relevant requester details  {$<$reputation$>$, (a unilateral form of \textit{ $<$sensitivity$>$}}); (2) Drawing insights from others' shared encounters with requesters  {\textit{ $<$collaboration$>$}}; and (3) Recognizing and interpreting information voids concerning prospective requesters  {\textit{ $<$articulation$>$}}. We delineate these three strategies as follows.

\subsubsection{Searching for Relevant Online Information} Every participant in our study affirmed that a key sousveillance strategy they employed involved actively seeking online information about their potential and existing requesters  {\textit{ $<$reputation$>$}}. Participants also reported using search engines such as Google [P15], inspecting their requesters' social media profiles for suspicious content [P9], looking up recent news about their requester's companies [P12], and searching for their requesters on career sites like LinkedIn [P8]. In addition, participants noted that they used this information not only to take into account the revenue of a company (to assist with wage negotiations) [P7]  {\textit{ $<$adversarial collaboration$>$},} but also to identify the requester's values or to discern whether their potential requester had a \say{\emph{trustworthy}} personality [P6, P8]  {\textit{ $<$trustworthiness$>$}}. 

\subsubsection{Learning from Previous Experiences}
In addition to researching online details about potential requesters, another key sousveillance strategy adopted by our participants included tracking the reported experiences of their fellow gig workers [P12, P15]  {\textit{ $<$collaboration$>$}}. Furthermore, they analyzed past data and encounters they had with particular requesters to inform their approach. Simply put, participants like P3 conceded that they \emph{\say{learn from [bad experiences] so that next time, [...] you’re more careful}}. Moreover, several gig workers [P3-4, P9, P12] emphasized their info-gathering strategies took place within online gig work communities and independent review sites, where fellow workers openly shared insights on particular requesters. This allowed workers to glean positive and negative experiences related to these requesters.
\subsubsection{Identifying and Interpreting Information Gaps} Despite the primary aim of sousveillance to collect information about one's requesters, our participants noted that \emph{not} being able to find information about a requester could be equally valuable [P3, P12, P15]. Therefore, a key sousveillance strategy workers embraced involved recognizing and deciphering data voids. For example, P15 recounted a situation where they were \emph{\say{trying to get information about the company, but couldn’t find any information on Google or anywhere.}} Because of this absence of information, P15 felt \emph{\say{like [they] couldn't trust [the requester] because it is weird that you can not find any type of information about a company on the Internet}}. Likewise, P12 firmly stated that if they saw no reviews on a requester, they \emph{\say{do not go ahead and apply}}. To these participants, an inability to find information about a requester indicated possible fraudulence  {\textit{ $<$articulation$>$}}. 

\subsection{Sousveillance Compels Gig Workers to Acknowledge Relationship Boundaries.} 
While discussing the potential benefits of sousveillance, our participants also considered how sousveillance might alter their relationships with gig platforms and requesters and make their position more difficult [P3, P9, P11-12]  {\textit{ $<$relationship$>$}}. According to P9, \emph{\say{there is some [requester information] that you don't want to know because then you [will] feel emotional blackmail}}  {(a negative form of \textit{ $<$empathy$>$}, \textit{$<$taking care of$>$})}. Although many gig workers expressed that they initially were wary of conducting sousveillance on their requesters [P1-3, P4, P6, P12, P14], in practice, they embraced sousveillance as a tool to support interactions with gig platforms and requesters wherever possible [P3, P8-9, P11]  {\textit{ $<$adversarial collaboration$>$  $<$reputation$>$}}. Nonetheless, as described directly below, we found that the act of conducting sousveillance often compelled workers to acknowledge and enforce work relationship boundaries, wherein workers upheld legal restrictions, respected requester preferences, and advocated for their own personal privacy  {\textit{ $<$respect$>$  $<$taking responsibility$>$}}.



\subsubsection{Conforming to Legal Restrictions} \label{legal}



Throughout our interviews, participants expressed how conducting sousveillance is entirely impossible or impractical if it does not strictly follow gig platform policies on data collection [P1-2]. In P2's words, \emph{\say{[if the] platform won’t allow you to do [sousveillance], it is [...] nearly impossible.}} Despite this legal obstacle, many participants depended on various forms of sousveillance to support their gig work and thereby deliberately chose to navigate these legal policies in order to conduct sousveillance [P4, P8-9, P12, P14-15]  {\textit{ $<$respect$>$}}. For example, according to those who regularly conducted sousveillance on their requesters like P8, \emph{\say{[sousveillance] is okay, as long as it’s transparent and approved.}} Thus, to workers like P8, P12, and P15, rather than view sousveillance as a defiant reaction against platform policies, they viewed it as a cooperative yet self-benefiting practice  {\textit{ $<$adversarial collaboration$>$  $<$respect$>$  $<$reputation$>$}}. Similarly, participants also emphasized the importance of abiding by regional and national data protection regulations when performing gig sousveillance [P1-2, P8, P12, P14]. Gig workers' awareness of legal restrictions pertaining to data collection and data privacy defined their sousveillance practices. For instance, P12 shared that when conducting sousveillance, they took into account \emph{\say{the [requesters'] country, [since] certain countries have rules in place like GDPR in the UK".}} In addition, because P12 was aware of GDPR's guidelines, they were therefore also cautious about collecting \emph{\say{[too] much information}} when conducting sousveillance. P14 similarly explained how conducting sousveillance could \emph{\say{feel like breaking the law}} because \emph{\say{it’s not appropriate to do unless [requesters] signed something [and] knew that [they were] doing [sousveillance].}} In the end, the necessity of practicing sousveillance compelled workers to be more attuned to legal considerations, leading to heightened awareness of the appropriate interactions and relationships that should exist between workers and their requesters  {(a unilateral form of \textit{ $<$mutual concern$>$} \textit{ $<$respect$>$  $<$taking responsibility$>$})}.

\subsubsection{Respecting Requester Boundaries}
During our interviews, we found that one of the strongest considerations that participants had when conducting sousveillance was the preferences of their requesters. As detailed further below, participants outlined that during the process of conducting sousveillance, they found it imperative to: 1) Adhere to requesters' terms  {\textit{ $<$attentiveness$>$  $<$caring about$>$  $<$sensitivity$>$}}; and 2) Align with requesters' work expectations  {\textit{ $<$competence$>$  $<$responsiveness$>$  $<$taking care of$>$}}. Several gig workers explained that requesters frequently enforce written contracts or non-disclosure agreements (NDAs) which can prevent them from conducting sousveillance altogether [P1, P12]  {\textit{ $<$respect$>$}}. P1 explained that although collecting data about requesters can be feasible on certain occasions, \emph{\say{There are many situations where we actually have to say, `Okay, we took a non-discretion act, and we are not supposed to talk about this'}}. These experiences indicate how the boundaries around sousveillance can vary by the requester. In essence, participants recognized that effective implementation of sousveillance hinged on their professional rapport with requesters {\textit{ $<$taking care of$>$  $<$taking responsibility$>$}}.

Participants also emphasized the importance of respecting requester-work expectations in order to maintain trusting and professional relationships [P2, P6, P8, P11]  {\textit{ $<$relationship$>$  $<$respect$>$}}. P6, a gig worker who operated as a ``virtual personal assistant'', speculated how a requester might react to a gig worker conducting sousveillance: \emph{\say{They might [not] think of [it] in a positive way. You know, like, [...] `Why are you requesting or collecting data about me when I’m [just] looking for an employee?'}.} Participants theorized that requesters strictly saw gig work as a business transaction and only employed workers to complete a desired task  {\textit{ $<$adversarial collaboration$>$}}  {(a limited form of \textit{$<$receiving care$>$})}. P9 pointed out that because requesters \emph{\say{don’t really care about [workers] [and] just want you to do the work}}, it can be difficult to obtain the requester's consent for conducting sousveillance  {\textit{ $<$relationship$>$}}. According to P3, P6, and P9, requesters only permitted workers to conduct sousveillance if it was deemed as \emph{\say{part of [their] job}}  {\textit{ $<$competence$>$  $<$taking care of$>$}}. Consequently, if a requester is hesitant about having sousveillance, P10 explained that workers should \emph{\say{feel safe to write a message to [requesters]}} and convey how conducting sousveillance would help the worker complete their job  {\textit{ $<$caring about$>$  $<$taking responsibility$>$}}. Participants considered that framing sousveillance as a benefit to their gigs would incentivize requesters to consent to the act of sousveillance  {\textit{ $<$taking care of$>$  $<$mutual concern$>$}}.

\subsubsection{Establishing and Maintaining Personal Boundaries} Gig workers revealed that practicing sousveillance occasionally led to an unintended gathering of information that was not immediately advantageous to them in a professional sense  {\textit{ $<$responsiveness$>$  $<$taking care of$>$}}. Instead, this information could potentially place them in emotionally challenging situations [P3, P9]. Thus, workers argued that identifying and maintaining personal boundaries is essential to safely conducting sousveillance  {\textit{ $<$relationship$>$}}. For example, P3 recounted an experience when they discovered unsettling information about their requester:
\emph{\say{[Too much information] would affect the way you view [requesters], and it could affect the way you deal with your work, even though that's not a part of your work. [...] You'll kind of start thinking, `Wait, should I stay here? Do I move to another job to find a better requester?' 
}}  {\textit{ $<$adversarial collaboration$>$  $<$reputation$>$}}

Likewise, P9 shared an experience where learning too much personal information about their requester made them feel emotionally obligated to help their requester with work beyond the scope of their contract. P9 explained how obtaining personal information about requesters can \emph{\say{make you have feelings for them [so] that you start feeling pity}} which can lead to emotional manipulation  {(\textit{caregiving}, a negative outcome of \textit{ $<$empathy$>$}, and a unilateral \textit{ $<$mutual concern$>$})}. In these two examples, although P3 and P9 successfully conducted sousveillance to learn more about their requesters  {\textit{ $<$responsibility$>$  $<$taking care of$>$}}, they struggled to emotionally process the information that they procured. Thus, to prevent themselves from inflicting unnecessary emotional harm, they recommended several strategies---such as avoiding requester social media accounts, selectively choosing sources from which to procure requester information, and limiting external communication with requesters---when conducting sousveillance  {\textit{ $<$relationship$>$}}.




In summary, in order to protect critical relationships while conducting sousveillance, our participants prioritized the welfare of themselves and their requesters by avoiding both the examination and collection of sensitive information. Moreover, these findings demonstrate how when conducting gig sousveillance, maintaining strict personal boundaries is imperative to minimizing unwanted violations, upholding requester expectations, and protecting personal well-being.

\subsection{Sousveillance Gives Freelancers a Competitive Edge in an Opaque Gig Economy}
Gig workers regarded this practice as a means to attain a competitive edge  {\textit{ $<$adversarial collaboration$>$}}, particularly when they were obliged to operate within platforms that exhibited a deficiency in transparency. In the following, we delve into the specifics of how they perceived sousveillance as a competitive advantage:

\subsubsection{Streamlining Communication with Requesters.}
Four participants reported using sousveillance to improve communication with their requesters [P8-9, P11-12]  {\textit{ $<$collaboration$>$}}. Workers recounted occasions when they encountered unresponsive or unreachable requesters, and the gig platforms remained opaque, failing to provide alternative means of contacting requesters or discerning their status  {(a negative form of \textit{ $<$coordination$>$})}. This prompted workers to engage in sousveillance to acquire their requesters' contact details outside of the gig platform {(\textit{$<$attentiveness$>$}, unilateral \textit{ $<$mutual concern$>$  $<$taking responsibility$>$  $<$taking care of$>$)}}. In P11's words, \emph{\say{getting the requester's phone [number] would help me as a gig worker to chat with a requester who is not active on the platform}}. Other workers like P9 and P12 kept track of their former requesters' personal contact information and professional social media accounts to be attentive to upcoming requests and maintain long-term communication with requesters  {\textit{ $<$attentiveness$>$}}. For example, P9 elaborated on their utilization of sousveillance to maintain a connection with a requester, grasp their preferences, offer aid with minor tasks, and establish an enduring communication channel: \emph{\say{If I speak with [my requesters] regularly, the chances that I’ll be forgotten is a little bit slow (P9)}} {\textit{ $<$coordination$>$  $<$reputation$>$  $<$taking care of$>$  $<$taking responsibility$>$}}. In this case, gig workers turned to sousveillance due to gig platforms' opaque nature, making requester updates hard to come by. Overall, gig workers shared their methodologies for overcoming the obscured communication hurdles presented by gig platforms, employing sousveillance as a proactive approach.

\subsubsection{Identifying Relevant Gigs.}
Similar to prior work \cite{gray2019ghost,toxtli2021quantifying}, participants described the process of searching and applying for gigs as tedious, time-intensive, and oftentimes unrewarding. Workers like P6 detailed experiences where the opacity of the gig platforms led them not to be well-informed about the details of a gig beforehand and thereafter \emph{\say{regret applying for [the gig]}} because they eventually realized that it did not correspond with their skills and interests  {\textit{ $<$competence$>$  $<$responsiveness$>$}}. Derived from these experiences, workers exhibited a strong tendency to carry out research on potential gigs they intended to apply for  {\textit{ $<$attentiveness$>$  $<$reputation$>$}}. They explained that they leveraged the insights gained through sousveillance to overcome the lack of transparency in their jobs and ensure \emph{\say{better matches}} that would ultimately enable them to \emph{\say{invest their time in more meaningful gigs}} [P2]  {\textit{$<$caregiving$>$ $<$taking responsibility$>$}}, \emph{\say{secure long-term gigs}} [P3], identify gigs that better \emph{\say{align with [their] values}} [P7]  {\textit{$<$relationship$>$ $<$sensitivity$>$}}, and allow them to \emph{\say{deliver more high-quality services [work]}} [P11]  {\textit{$<$caregiving$>$ $<$competence$>$ $<$taking responsibility$>$}}.

\subsubsection{Reducing the Threat of Scams.}
 Over half of our participants [P2-3, P8, P10-15] indicated they adopted sousveillance techniques to proactively detect and thwart scams prevalent on gig platforms. These scams involved cases where malicious individuals posed as requesters or posted fake gigs  {(\textit{adversarial collaboration}, inverse \textit{$<$mutual concern$>$}, \textit{trustworthiness})}, often exploiting the absence of transparency on gig platforms, which rendered the detection of fake requesters or fraudulent tasks difficult for workers. According to P15, gig workers find themselves inadequately supported by platforms to shield against scams, rendering them particularly vulnerable to fraudulent activities. Consequently, they advocated for fellow workers to adopt sousveillance as a proactive measure in safeguarding themselves against scammers: \emph{\say{Freelancers usually don't enjoy many protections. Because of that, we have to look after ourselves, and so it is very important that every freelance worker gathers as much information as they can about any requester or a platform...}}  {\textit{$<$collaboration$>$ $<$reputation$>$}}. Like P15, P12 relied on sousveillance to conduct preliminary background checks on potential requesters to mitigate the threat of getting scammed. P12 explained that, \emph{\say{If there is something [about a requester] which is publicly available, I do my part of [conducting] research to make sure that I am not being scammed}}. Other workers highlighted their practice of scrutinizing individual requester details as a preemptive strategy to steer clear of scams, including a focus on verified payment methods and listings [P3, P7, P15], gig workers reviews [P8, P12], or backing by reputable companies [P12]. For instance, P12 shared that \emph{\say{if there’s a requester who has no payment method verified, or they don’t have previous reviews, I do not go ahead and apply}}. Ultimately, given the recurring instances of fraudulent activities faced by gig workers \cite{ravenelle2022good}, and the limited transparency offered by gig platforms to easily identify the scammers, several of our participants deemed sousveillance an indispensable and pragmatic tool to detect and counteract scams effectively  {\textit{$<$adversarial collaboration$>$ $<$reputation$>$}}.

\subsection{Reservations Among Gig Workers Regarding Sousveillance Practices.}
Participants also expressed reservations about conducting sousveillance. Many participants reported that they hesitated to conduct sousveillance based on two key assessments: 1) the overhead cost of conducting sousveillance  {\textit{$<$attentiveness$>$}}; and 2) the risk of damaging relationships with requesters  {\textit{$<$relationship$>$}}. Next, we delve deeper into workers' pre-sousveillance appraisals.

\subsubsection{Weighing Necessary Investment Costs with Potential Outcomes}
One of the most important things that workers considered when choosing whether to conduct sousveillance was the amount of effort necessary to reap the benefits of sousveillance. Gig workers expressed how conducting sousveillance felt like an additional burden on top of having to constantly apply for new gigs, send updates to current requesters  {\textit{$<$articulation$>$ $<$coordination$>$}}, and advertise themselves to prospective requesters [P11-12, P14]  {\textit{$<$reputation$>$}}. For instance, P14 explained that aside from conducting simple background checks of requesters, they felt like they did not need to conduct sousveillance because they, \emph{\say{didn't have the tools}} and \emph{\say{[were] already busy enough}}. In contrast, other workers [P3, P8, P15] viewed sousveillance as a tool to prevent them from wasting precious time with non-cooperative or distasteful requesters. P8 declared that through sousveillance, they \emph{\say{limit wasting time with a requester which is not worth it, meaning if [they] see a problematic requester, [they] don’t even go ahead [and apply for the gig]}}  {\textit{$<$adversarial collaboration$>$}}. Altogether, we saw that many participants grappled with identifying which circumstances most benefit from conducting sousveillance.


\subsubsection{Balancing Freelancer-Requester Dynamics} When choosing sousveillance (or having reservations), participants weighed its potential to improve work conditions against possible requester relationship damage  {\textit{$<$relationship$>$ $<$respect$>$}}. They feared backlash, harming reputation, and legal consequences [P2, P8-9, P12, P14]  {\textit{$<$reputation$>$}}. To begin with, since many workers expressed the importance of maintaining a healthy working relationship with their requesters, they also worried that conducting sousveillance on their requesters could be misinterpreted as mistrust [P2-4, P6, P11-12]. For instance, P2 mentioned that, to a requester, conducting sousveillance could seem as though a worker were saying \emph{\say{I need to keep myself protected, because I do not trust you}}  {\textit{$<$adversarial collaboration$>$}}. Moreover, participants also mentioned that there is significant liability involved with conducting sousveillance. In particular, P9 and P11 also expressed that if confidential information about one of their requesters was accidentally leaked, they could be accused of the data leak, even if they weren't directly responsible for the incident  {\textit{$<$caregiving$>$ $<$reputation$>$ $<$taking responsibility$>$}}. Consequently, participants highlighted the need to ensure both mutual transparency and benefit when conducting sousveillance [P2, P4, P6-9]. For example, P4 revealed that when they \emph{\say{collect information on the requester, it is best that the requester knows about it, to avoid any further issue}}; P8 emphasized how the goal of sousveillance should be to create \emph{\say{added value for [workers]}} as well as \emph{\say{added value for requesters}}; and P9 similarly argued that sousveillance should help to embrace the \emph{\say{symbiotic relationship}} between workers and requesters. Altogether, our participants stressed the importance of conducting sousveillance in a way that benefited and supported both workers and their requesters  {\textit{$<$mutual concern$>$}}.

\section{Results: Co-Design Activity}
Our  {co-design} analysis revealed three key themes for future sousveillance technology design for gig workers. These themes emphasize the need for tools that can: (1) Ease the emotional strain of gig work  {(inverse unilaterial  {\textit{$<$empathy$>$}}, \textit{$<$taking care of$>$})}; (2) Facilitate the acquisition of preferred gigs  {\textit{$<$adversarial collaboration$>$}}; (3) Monitor and manage the invisible labor demands of gig work \cite{toxtli2021quantifying,gray2019ghost}  {\textit{$<$articulation$>$}}.

\subsection{Searching for Legitimate Gigs}



Much like in our interviews, participants emphasized the challenge of identifying scams and scammers during our  {co-design} activity  {(\textit{$<$adversarial collaboration$>$}, inverse \textit{$<$mutual concern$>$, $<$trustworthiness$>$}}. They emphasized their vision of future sousveillance tools assisting in this critical identification process. Certain workers envisioned these tools offering transparency regarding a requester's company, prior gigs hired, and reviews from fellow gig workers [P1, P4, P7-8, P12, P14-15]  {\textit{$<$reputation$>$\\ $<$trustworthiness$>$}}. Notably, this design concept bears resemblance to existing sousveillance tools like Turkopticon, tailored for the Amazon Mechanical Turk crowdsourcing platform \cite{irani2013turkopticon}. However, other workers, like P11, expressed how they envisioned future sousveillance integrating additional information. They believed this type of information is needed to disprove fraudulence and noted how \say{\emph{[there is no way to know that reviews are genuine and are not paid for}}. Moreover, P14---who was underpaid by a five-star-rated requester---described how requester reviews are often misleading because \say{\emph{there's a stigma if you don't give someone a five-star review}}. To these workers, robust sousveillance should include not only the act of obtaining information about one's requesters but also evaluating the integrity of that information. 


\subsection{Alleviating the Emotional Toll of Gig Work} 
While most participants primarily focused on suggestions for sousveillance tools supporting their gig work, they also stressed the significance of creating sousveillance tools that gather data to enhance worker well-being  {(unilateral, self-focused \textit{$<$mutual concern$>$} and \textit{$<$caregiving$>$})}. For example, many workers described habits where they would \say{\emph{stare at a screen for long hours}} [P4], \say{\emph{forget to drink water}} [P6], work \say{\emph{16 hour days}} [P14], \say{\emph{forget to take breaks}} [P3], \say{\emph{and work nonstop}} [P3]. To combat these unhealthy work practices, these participants suggested the use of surveillance tools that could promote workers' mental, emotional, and physical health. For instance, P3 envisioned a tool that could give workers \say{\emph{motivational pushes}} and \say{\emph{reminders to be kind [to] yourself}}, whereas P15 suggested implementing reminders to \say{\emph{take a lunch break}} and to \say{\emph{take a deep breath and stretch}}. Moreover, workers P1, P6, and P16 further specified that such a tool should ideally accommodate to different workers' varying work styles and individual health needs. P6 explained that, \say{\emph{[D]uring our shift, we're mostly supposed to stay near our laptops or desktop, so going for a walk suddenly is not really the best recommendation}}  {\textit{$<$taking responsibility$>$}}. Likewise, P16 suggested that a worker might need different types of suggestions \say{\emph{if [they] have a particular or special kind of ailment}}. 



\subsection{Measuring and Managing Invisible Labor} 
Several participants had used self-monitoring tools (e.g., Toggl or Traq) in response to requesters' demands [P5, P8, P16]  {\textit{$<$taking care of$>$ $<$taking responsibility$>$ $<$responsiveness$>$}}. Our participants expressed the desire to repurpose these surveillance tools to instead analyze and improve their performance for their own personal benefit [P3-4, P7-9, P11-12, P14]. For example, P2 envsioned that such a tool \say{\emph{could make [a worker be] very organized, [know] how much time [they're] spending on which task, and prioritize tasks}}  {\textit{$<$caregiving$>$ $<$care receiving$>$ $<$taking responsibility$>$}}. Other workers also brainstormed self-surveillance tools that not only visualized workers' logged work hours but also offered suggestions on how to improve their productivity. In the design mockups intended to inspire workers' visions of future sousveillance tools, we incorporated details about workers' earnings and unpaid labor (invisible labor), particularly emphasizing instances when earnings fell below the minimum wage  {\textit{$<$adversarial collaboration$>$ $<$articulation$>$}}. This feature was included because previous research on sousveillance tools for crowdworkers has primarily centered around this variable \cite{toxtli2021quantifying,li2022all}. However, when presented with sousveillance tools that integrated this type of feature, several workers expressed concerns [P9, 11, 14-15]. In particular, P15 explained the potential emotional harm that tracking wages and invisible labor could cause, stating that, \say{\emph{I feel like I'd get depressed seeing how much unpaid work I do}}. Likewise, P9 also commented that because it can be difficult for gig workers to find high-paying gigs, such a tool could feel \say{\emph{hopeless}} and \say{\emph{condescending.}}



\section{Recommendations for Designing Worker-Centered Gig Sousveillance Systems}
 {Based on our interviews and co-design activities, we formulated four design recommendations for gig sousveillance tools, each aligned with at least care ethics principle. These recommendations are illustrated in Figure \ref{fig:design-recs}}. 

 {Note that in designing tools for gig workers, like these sousveillance technologies, it is vital to recognize they often emerge in response to the lack of care ethics in gig platform design \cite{savage2021research,toxtli2023designing}. In particular, gig markets are usually structured to prioritize short-term, flexible jobs and efficiency over workers' well-being \cite{Anicich_2022}, often also neglecting crucial aspects of care ethics like interpersonal relationships and workers' (and even requesters') individual needs \cite{gray2019ghost}. Gig platforms also often have designs that restrict worker support and protection in vital areas such as their safety, financial security, and professional development \cite{kessler2018gigged,woodcock2020technology,mexi2019gig}. These limitations are in contrast to the fundamental principles of care ethics \cite{held2006ethics,tronto2020moral}, which emphasize the importance of safeguarding these critical aspects for workers \cite{held2006ethics,ley2023care}. This underscores the necessity for a more supportive and caring strategy on gig platforms—a void that could be effectively filled by worker-centric tools, such as gig sousveillance technologies.}

\begin{figure}
    \centering
    \includegraphics[width=400pt]{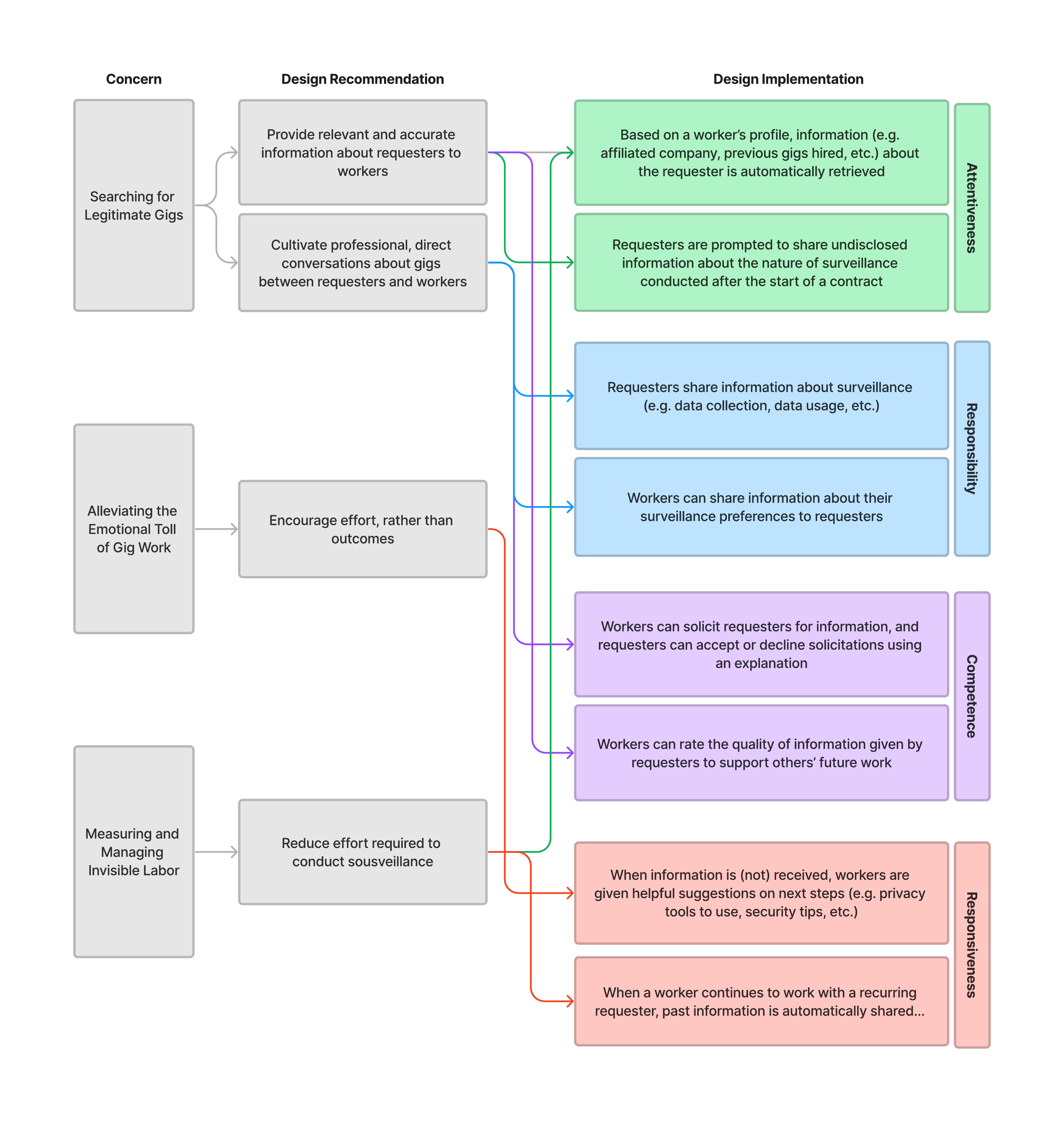}
    \caption{Concerns, design recommendations, and design implementations based on interviews and  {co-design} activities with participants. All of the proposed design implementations also correspond with each of the four care ethics principles. We offer a total of four design recommendations and eight suggested design implementations.}
    \label{fig:design-recs}
\end{figure}


\subsection{ {Design Recommendation: Create Sousveillance Technologies Balancing Data Accuracy, Attentiveness, and Data Competence for Workers and Requesters.}} 
 {Participants emphasized challenges in identifying genuine gigs and suggested using sousveillance to evaluate requester trustworthiness  {\textit{$<$trustworthiness$>$}}. They stressed the importance of being able to retrieve accurate information about requesters  {\textit{$<$reputation$>$}}. Many participants dedicated considerable time to scrutinizing requesters' social media profiles, particularly on LinkedIn, and employed tools like TurkOpticon \cite{irani2013turkopticon} to gauge other workers' experiences and opinions about these requesters  {\textit{$<$collaboration$>$}}. Their objective was to have rich and accurate data about their requesters  {\textit{$<$attentiveness$>$}}. Preferences about the type of data they desired varied: some sought details about a requester's company size or revenue, while others focused on requester profiles. Identifying data voids \cite{golebiewski2019data,flores2022datavoidant} was also seen as a way to verify requester authenticity.}  {Based on this, we recommend designers create sousveillance systems that provide comprehensive and accurate requester information to workers  {\textit{$<$adversarial collaboration$>$ $<$reputation$>$ $<$trustworthiness$>$}}. Such designs could include a ``checklist" outlining what information about a requester is available or unavailable, based on the preferences of a worker  {\textit{$<$attentiveness$>$}}. By obtaining indications of both readily available and unavailable information, workers could develop an immediate baseline impression of a requester's authenticity, based on information that workers individually view as important before starting a gig.}  {However, note that these designs can also be enhanced by aligning with the principles of care ethics \cite{ley2023care,ripamonti2021care}. In particular, we believe such designs should also embody ``Attentiveness" where there is a conscientious focus on the needs, feelings, and contexts of all involved parties \cite{ley2023care,noddings1995caring}  {\textit{$<$attentiveness$>$}}. This approach would thus require not only catering to workers' information needs to discern authentic opportunities, but also consider requesters' perspectives \cite{tronto2020moral,held2006ethics}  {\textit{$<$relationship$>$}}. Attentiveness demands a deliberate and considerate approach. Therefore, it is crucial to ensure that requesters' consent and comfort are also prioritized. This could involve designing interfaces that transparently communicate the rationale behind monitoring requesters  {\textit{$<$mutual concern$>$}}, particularly emphasizing the protection and welfare of workers.}

 {The design of these sousveillance systems can also be further enhanced by integrating the care ethics principle of ``competence" \cite{ley2023care,ripamonti2021care}  {\textit{$<$competence$>$}}. Here, designers should aim to ensure that sousveillance systems not only monitor effectively but also operate with an awareness of the level of skill and proficiency of workers and requesters. The principle of competence emphasizes the need for sousveillance systems that are not only vigilant but also adept in their functionality. In this case, sousveillance systems should not only provide comprehensive lists covering the information about  requesters  {\textit{$<$reputation$>$}}, but also be attuned to the varied needs and backgrounds of workers. This could involve creating a dynamic checklist that adapts to the worker's experience level, offering culturally sensitive insights, as well as  incorporating feedback mechanisms for continual improvement  {\textit{$<$sensitivity$>$}}. Prioritizing competence also involves a careful balance between gathering comprehensive information from requesters and respecting their privacy \cite{hofstede1984culture,li2017cross}. This approach is particularly important for new requesters who may not have detailed information about themselves, and for those from cultures where sharing personal details is less common \cite{trepte2017cross,hofstede2009geert}. Integrating tools for competence support, such as tutorials or case studies, can further empower workers in making informed decisions \cite{rabin2021care}. Additionally, ensuring the system is accessible and inclusive to all, regardless of their digital literacy or cultural background, is essential \cite{seton2023care}. This approach can not only enhance the functionality of sousveillance systems, but can also foster an environment of informed decision-making, cultural understanding, and respect for diversity, which are integral to the care ethics principle of competence \cite{gunaratnam2008competence,hamington2011care,noddings2015care}.}

\subsection{ {Design Recommendation: Create Interfaces that Enable Professional Dialogue on Sousveillance.}}
 {In each session, participants affirmed that the crux of sousveillance was its impact on their relationships and interactions with requesters  {\textit{$<$relationship$>$}}. Many of our participants advocated that in order to conduct sousveillance, it must be viewed as ``professional", ``permissible", and ``transparent" to their requesters. Consequently, several participants explained that they often would remind their requesters of the reasons and benefits behind conducting sousveillance.} 

 {Based on these assertions, we propose that gig sousveillance systems should uphold mutual understanding and mutual benefit for both requesters and workers  {\textit{$<$mutual concern$>$}}. In particular, we recommend that a gig sousveillance system should facilitate conversations between workers and requesters  {\textit{$<$collaboration$>$ $<$relationship$>$ $<$respect$>$}} to ensure that 1) requesters understand the direct benefits of sousveillance (e.g., increased productivity, less time spent on a gig, less time spent discussing a contract), 2) requesters consent to sousveillance, 3) workers act lawfully, and that ultimately 4) professional and collaborative work relationships between workers and requesters are nurtured through open communication. We imagine that these design recommendations could involve using chatbots as intermediaries in labor interactions. Chatbots could obtain consent and clarify contexts for requesters, easing the awkwardness often present in communications, especially when asking for input from those in power. This approach, supported by prior work \cite{toxtli2018understanding}, can enhance comfort in these exchanges}.  {Note that these design recommendations are consistent with care ethics principles, focusing on ``attentiveness'' and ``responsibility'' \cite{ley2023care,ripamonti2021care}  {\textit{$<$attentiveness$>$ $<$taking responsibility$>$}}. Especially, as these design recommendations prioritize being attentive and helping all parties understand sousveillance benefits, while also being responsible and securing requester consent, as well as ensuring lawful worker behavior \cite{klaver2011attentiveness,randall2018values}. This approach respects both parties' autonomy, fostering responsible and informed engagement in professional relationships \cite{randall2018values}  {\textit{$<$relationships$>$ $<$respect$>$}}. Using chatbots can facilitate respectful communication in power-imbalanced worker-requester interactions, emphasizing ethical practices.  {\textit{$<$mutual concern$>$}}.}

\subsection{ {Design Recommendation: Create Responsive Sousveillance Technologies Focused on Emotional Well-being Instead of Uncontrollable Work Outcomes.}} 

 {In our sessions, workers noted that while sousveillance can empower them prior to working with a requester  {\textit{$<$adversarial collaboration$>$}}, it may also cause emotional stress. Learning unsettling facts about requesters or realizing through sousveillance that they are underpaid can lead to feelings of demotivation and helplessness.} 

 {Building on these insights, we advocate for sousveillance systems to offer encouraging and constructive followup to workers regarding their data discoveries with sousveillance. This approach should also emphasize the importance of valuing workers' contributions and efforts  {\textit{$<$mutual concern$>$}}. To enhance worker empowerment and support, these systems should concentrate on aspects within workers' direct influence  {\textit{$<$competence$>$ $<$taking responsibility$>$},} rather than factors beyond their control, such as low wages. This approach could reduce feelings of depression, helplessness, or powerlessness by fostering a sense of agency and involvement in their work environment \cite{zhou2021does,mccormack2013managing}. Note that this design recommendation contrasts with typical sousveillance systems that expose gig market realities \cite{toxtli2021quantifying,calacci2022}. These systems, while potentially beneficial for activism, risk causing workers' to feel helpless. The key may lie in striking a balance between transparency and empowerment \cite{gunawan2023effect,chavez2022liderazgo}. A potential solution is to offer workers follow-up support that includes guidance on practical steps for positive change \cite{guo2022relationship}}.  {Note that this design recommendation can be further improved by integrating the responsiveness aspect of care ethics \cite{hamington2019integrating,reynolds2016infinite}  {\textit{$<$responsiveness$>$}}. Accordingly, sousveillance systems should go beyond just offering positive and constructive feedback. They should proactively address and respond to workers' specific needs and concerns when interacting with sousveillance data \cite{randall2018values}. This approach would involve a deeper engagement with workers' experiences, especially acknowledging their efforts and contributions in a meaningful way  {\textit{$<$respect$>$}}. By demonstrating responsiveness to the emotional needs of workers, such as addressing depression and feelings of powerlessness, sousveillance systems could contribute to creating a more nurturing and empowering workplace \cite{cartwright1997managing,shields2006stress}  {\textit{$<$mutual concern$>$}}. In this responsive environment, workers would feel valued and acknowledged, with their well-being being as prioritized as their productivity \cite{hanley2020eyes}.}

\subsection{ {Design Recommendation: Simplify Sousveillance for Gig Workers via Predictive, Attentive, and Responsive Sousveillance Mechanisms.}} 

 {Participants expressed concerns about having to invest too much time in sousveillance and how that time investment would impact their workloads. To alleviate this burden from gig workers, we propose two key design approaches: 1) implementing predictive sousveillance mechanisms that automatically offer insights based on a worker's preferences and 2) prompting requesters to share information about themselves and their gigs  {\textit{$<$reputation$>$}}. In particular, by diverting the responsibilities and actions of sousveillance, workers can have greater pace of mind and focus on other responsibilities. We see value in automating aspects of the sousveillance process, as prior research has begun to investigate \cite{toxtli2021quantifying,hara2018data}. Ultimately, these proactive measures can not only reduce the mental strain and effort required for sousveillance but also ensure a more harmonious and mutually beneficial relationship between gig workers and requesters  {\textit{$<$collaboration$>$ $<$relationship$>$ $<$respect$>$}}.}  {Incorporating the care ethics principle of responsiveness into these design recommendations could significantly enhance their effectiveness \cite{hamington2019integrating,reynolds2016infinite}  {\textit{$<$responsiveness$>$}}. Responsiveness, which emphasizes understanding and addressing individual needs, suggests the potential for customizing predictive sousveillance tools to align with each worker's specific preferences. This tailored approach would streamline the process, making it more user-friendly for workers and providing more relevant, supportive insights. We can also imagine that these designs could be further enhanced by embedding the care principle of attentiveness into the predictive sousveillance mechanisms  {\textit{$<$attentiveness$>$}}. Through this, these sousveillance tools could become more than just automated systems; they could evolve into intuitive platforms that adapt and respond to the changing circumstances and preferences of individual workers. This means the system would not only offer insights based on a worker’s stated preferences but also learn from their interactions and feedback, continuously refining its support. Conversely, responsiveness in care ethics advocates for fostering mutual respect and understanding  {\textit{$<$mutual concern$>$ $<$respect$>$ $<$taking responsibility$>$}}. There is thus likely also value in designing sousveillance interfaces that educate requesters on the benefits of transparency, and with it potentially nurturing a more empathetic and collaborative work environment  {\textit{$<$collaboration$>$}}. Such an approach could transform worker-requester interactions from mere transactions to relationships characterized by mutual care and respect  {\textit{$<$relationship$>$ $<$respect$>$}}. By integrating responsiveness into these designs, we can create a more supportive and understanding framework for both workers and requesters. Furthermore, when prompting requesters to share information about themselves and their gigs, the care ethics principle of attentiveness could also be integrated  {\textit{$<$attentiveness$>$ $<$reputation$>$}}. This might result in ensuring that these requester prompts are relevant, timely, and sensitive to the context in which gig workers operate. This could involve intelligently determining the most opportune moments to request information, thereby minimizing disruption and respecting the worker's time and mental load. By weaving attentiveness into these design strategies, the systems become more than just functional tools; they transform into empathetic allies that understand and adapt to the dynamic needs of gig workers. This approach not only eases the mental burden associated with sousveillance but also fosters a more supportive, efficient, and mutually beneficial environment for both gig workers and requesters. Ultimately, these attentive design enhancements can lead to a significant reduction in the mental strain and effort required for effective sousveillance, thereby improving the overall experience for all parties involved.}

\section{Discussion}
 {This discussion synthesizes key insights from our study and explores the application of care ethics, collaboration, and sousveillance technologies.}
\subsection{Lessons from Conducting Co-Design Sessions with Workers}
While there exists limited research which has employed  {co-design} methods to create user-centered tools for gig workers, there is a growing body of interest in applying HCD methods to gather critical perspectives of the unique needs and motivations of diverse, global gig workers \cite{zhang2022,lee2021,hsieh}. Prior work has noted the strengths of using  {co-design} methods to explore workers' issues. For example, Lee et. al. proposes that the use of participatory methods stimulates worker empowerment, the concept that workers can ``engage in relevant decision-making in the workplace \cite{lee2021}. Similarly, according to Huybrechts et. al., who cites the historical connection of  {co-design} to ``labour unions in emancipating workers at the workplace",  {co-design} serves as a form of ``political activism" wherein workers---who usually work in isolation---have the opportunity to collaboratively share and contribute their opinions \cite{huybrecht_codesign,huybrechts_reactivate}. As such,  {co-design} offers additional benefits in uplifting and advancing worker voices. In our study, we found that harnessing HCD methods allowed us not only to identify further social, legal, and emotional barriers to conducting sousveillance, but also to realize opportunities where we could prioritize the well-being of workers in gig sousveillance systems. In particular, many participants shared how conducting sousveillance could not only lead to possible conflict with platforms, but also potentially damage relationships with certain clients. Thus, through our  {co-design} activity, we gathered workers' feedback and vision for sousveillance systems, allowing us to recognize design recommendations grounded in workers' beliefs and experiences. In addition, three methodological supports underpinned our  {co-design} activity: 1) contextualizing our  {co-design} activity with a semi-structured interview and 2) providing participants with an informational workbook, and 3) walking participants through mockups as a design probe. Together, these techniques enhanced the focus and comprehension of our participants, many of whom spoke English as a second language. Ultimately, by incorporating HCD methods, we can ensure that sousveillance empowers workers to navigate the gig economy with greater control over their data and privacy without sacrificing their safety or comfort.

\color{black}
\subsection{Ethics of Care and Collaboration Reconsidered}
\color{black}
 {At first glance, gig work seems inherently aligned with care ethics. For example, care ethics emphasizes the importance of responding to individual needs and situations \cite{reynolds2016infinite,ley2023care}. Gig work offers workers the flexibility to choose when, where, and how much they work \cite{jarrahi2020platformic,mulcahy2016gig,katz2019rise}. Similarly, care ethics values diversity and inclusivity \cite{tronto2013caring}. Gig work also often serves as an inclusive employment alternative \cite{woodcock2020technology}, accommodating a broad spectrum of individuals, especially those who might face challenges or marginalization in conventional job markets \cite{kessler2018gigged,strauss2013temporary}. Furthermore, gig work usually benefits from its diverse and inclusive workforce \cite{mcgovern2017thriving}, especially because it leverages varied perspectives and experiences for complex and extensive data gathering \cite{malone2018superminds}.}

 {Surprisingly, our study uncovered a paradox in the realm of gig work as it relates to care ethics: despite the inherent potential for strong connections to care ethics in gig work, we observed that gig workers did not demonstrate significant caregiving traits, neither among themselves nor towards requesters. Note that our interviews and co-design sessions did reveal that workers demonstrated qualities such as attentiveness, responsiveness, responsibility-taking, and trustworthiness towards both their fellow workers and requesters (qualities which align with Tronto's initial phases of care,  where individuals become aware of unmet needs in those around them, identify how to address these needs, and decide to take appropriate action). Nevertheless, it proved intriguing to note that workers did not give much thought to the advanced stages of caregiving as outlined by Tronto \cite{tronto2013}. These stages encompass concrete caregiving, involving the actual provision of care with direct interaction with the individual in need, and care receiving, which involves recognizing, receiving, and acknowledging the care being provided. Overall, during our interactions with workers, a noticeable lack of empathy and mutual concern emerged, especially in their interactions with their fellow workers and requesters, unless they perceived them as potential liabilities. It is also important to mention that, as reported by gig workers, requesters did not exhibit any of the caregiving stages outlined by Tronto \cite{tronto2013caring}.
In summary, despite the potential for a culture of care in gig work, there appears to be a limited range of caring behaviors among workers and none in requesters.}  {The limited caregiving in gig work could be attributed to platform policies and designs that discourage collaboration \cite{gray2019ghost,christiaens2022digital}. Gig platforms, by lacking collaborative features, not only fail to uphold care ethics but also impede collaborations that could enhance these ethics, leading to a cycle where the absence of support further diminishes care ethics in the workplace. Suchman emphasized designing technology to aid, not dictate, human actions \cite{suchman1987plans}. The absence of collaborative interfaces in gig platforms likely enforces a rigid, machine-focused work approach, neglecting the necessity for adaptable, human-centric interactions \cite{wood2018workers,vallas2020platforms}. The rigid design of gig platforms contrasts sharply with the dynamic, context-driven nature of human work \cite{suchman1987plans,de2015rise,kalleberg2009precarious}, leading workers to create their own solutions and workarounds. This underscores the importance of designing flexible, human-centered technology that doesn't restrict communication among workers. We believe that the implementation of worker sousveillance technologies has the potential to effectively bridge these gaps and significantly enhance collaboration. Prior research indicates that shared data resources can foster communities and collaborations \cite{sawyer2014digital,barley1986technology}. By providing workers with tools to share workplace data, we could also encourage community building and collaborative efforts.}

 {On the other hand, it is crucial to recognize that the inherently competitive nature of these platforms often creates a scenario where workers are pitted against one another \cite{nilsen2023health,mckercher2013precarious}, further discouraging caregiving \cite{kessler2018gigged,rosenblat2018uberland,wu2022happy}. Unlike traditional workplaces with HR departments or support programs, gig platforms lack structures to facilitate care, making it difficult for workers to access support or even identify who needs care. The absence of communication channels also limits information and emotional support sharing. Without a collaborative environment, gig workers may also not develop a strong collective identity, essential for building empathy and promoting caregiving. Moreover, gig platforms typically do not provide incentives for giving care to others. Without incentives for caregiving behaviors, workers are less motivated to engage in such activities, especially if the platform even penalizes collaboration \cite{gray2019ghost}}.  {Drawing on insights from social theorists Tronto, Held, and Engster \cite{engster2007heart,held2006ethics}, as well as the second sensitizing concept of collaboration \cite{boone2023data}, we propose a novel approach to address the lack of caregiving in gig work: sousveillance technologies for collaborative caregiving. Envisioned as worker-owned tools, these sousveillance technologies would empower workers to actively monitor and enhance not only their own welfare but also that of their peers. To encourage participation, we suggest incentivizing engagement in various caregiving phases, including identifying needs, devising solutions, providing care, and acknowledging the care received \cite{tronto2013}. This process will likely not only facilitate care but also cultivates a sense of solidarity among workers.}  {To further enhance this concept, we envsision creating advanced sousveillance tools specifically designed for collaborative purposes, with a strong focus on facilitating collective action. These tools would not only support individual caregiving tasks but also seamlessly integrate functionalities that enable workers to unite and coordinate their efforts, amplifying their collective voice and impact.}  {Prior research on sousveillance technologies for workers has predominantly centered on quantifying workers' conditions to expose their harsh reality \cite{toxtli2021quantifying,hara2018data,griffith2022,li2022all}. However, our interviews revealed the negative impact of presenting such demoralizing data to workers. Consequently, we advocate for a shift in focus towards sousveillance technologies that can quantify the value of workers' labor. This could significantly empower workers, aiding them in effectively planning labor strikes and other collective actions \cite{calacci2023building}. By doing so, sousveillance becomes a potent instrument for enhancing worker empowerment and advocacy.}

\subsection{The Future of Sousveillance Technologies for Gig Workers}
As the landscape of the modern workplace undergoes swift transformations, the integration of sousveillance technologies tailored for workers presents a promising avenue for the creation of new data-centric systems \cite{jarrahi2021flexible,alvarez2022design,de2023independiente,de2023independiente2}. These systems are poised to empower workers by providing them with the insights necessary to make well-informed choices \cite{savage2020becoming,hanrahan2021expertise}, augment their productivity and well-being \cite{schlicher2021flexible}, and advocate for improved working conditions \cite{savage2016botivist,salehi2015}. Particularly, this data can play a pivotal role in helping workers discern recurring issues, rather than perceiving them as isolated incidents, thus highlighting systemic challenges. Furthermore, these tools could be synergized with platforms utilized by journalists, thereby enabling the media to shed light on the adversities faced by workers through evidence-based reporting. Envisioning further, such data-driven frameworks could significantly benefit gig workers by facilitating skill enhancement and boosting professional visibility \cite{chiang2018crowd,sarasua2015crowd,kasunic2019crowd,rivera2021want}. This could be achieved by offering personalized training recommendations and chronicling their career advancements. For the maximization of these systems' effectiveness, it is imperative to prioritize the development of solutions centered around the workers' needs \cite{zhang2022,li2022bottom}. This includes devising strategies for the restitution of data to the workers, thereby empowering them to leverage this information in sculpting the futures they aspire to. This approach ensures that the technology serves as a tool for empowerment rather than surveillance, fostering a more equitable and responsive workplace environment.

\subsubsection*{Limitations and Future Work}
Our work investigates gig workers' perceptions, attitudes, practices, and expectations of gig work sousveillance. Although our sample size includes male, female, and non-binary gig workers from across three continents, our results may not generalize to the opinions and experiences of all gig workers around the world. Moreover, because there exists an innumerable amount of gig work skills, specialties, and industries, our results may not reflect the experiences of gig workers who work outside of the areas that our research examines. However, we hope that our paper encourages future work that will more closely investigate the perceptions, attitudes, practices, and expectations of gig work sousveillance in specific subdomains of gig work. We also acknowledge that the topics of surveillance and sousveillance are especially sensitive to many gig workers. Because we aimed to write interview questions that would study the topic of gig work sousveillance without broaching overly sensitive topics that could potentially endanger gig workers (e.g., breaking non-disclosure agreements), our results are possibly impacted by response bias. Finally, while our  {co-design} activity sheds light on best design practices for designing sousveillance tools for gig workers, future work should further test, implement, and iterate upon our designs to better account for the situations and challenges that workers typically face. We hope that our research inspires future work in   {co-design}ing gig sousveillance tools that can enact real change in the daily experiences of gig workers.

\section{Conclusion}
We interviewed 16 gig workers to understand their perspectives on gig work sousveillance. Many already use sousveillance to assess requesters' needs but have concerns about its cost and potential conflicts. We propose a gig worker-centered tool for gathering requester information while promoting well-being and relationships. Our research blends sousveillance and an ethics of care framework, aiming to inspire human-centered sousveillance tool development prioritizing worker welfare.\\

{\bf Acknowledgements}
This work was partially supported by the NSF Grant FW-HTF-19541, Northeastern University's Office of Undergraduate  Research and Fellowships, especially for awarding Maya with the AJC Merit Research Scholarship and Peak Trail-Blazer Award. We also want to thank the Feminist AI Network, the A Plus Alliance for Inclusive Algorithms.

\bibliographystyle{ci-format}
\bibliography{ci2018-sample-bibfile}

\end{document}